# Probing single unit-cell resolved electronic structure modulations in oxide superlattices with standing-wave photoemission

W. Yang,[1,2] R. U. Chandrasena,[1,2] M. Gu,[3] R. M. S. dos Reis,[4] E. J. Moon,[5] Arian Arab,[1,2] M.-A. Husanu,[6,7] J. Ciston,[4] V. N. Strocov,[6] J. M. Rondinelli,[3] S. J. May,[5] A. X. Gray[1,2,*]

[1] *Department of Physics, Temple University, Philadelphia, Pennsylvania 19122, USA*

[2] *Temple Materials Institute, Temple University, Philadelphia, Pennsylvania 19122, USA*

[3] *Department of Materials Science and Engineering, Northwestern University, Evanston, Illinois 60208, USA*

[4] *National Center for Electron Microscopy, Molecular Foundry, Lawrence Berkeley National Laboratory, Berkeley, California 94720, USA*

[5] *Department of Materials Science and Engineering, Drexel University, Philadelphia, Pennsylvania 19104, USA*

[6] *Swiss Light Source, Paul Scherrer Institute, CH-5232 Villigen, Switzerland*

[7] *National Institute of Materials Physics, Atomistilor 405A, Magurele 077125, Romania*

*axgray@temple.edu

# Abstract

Control of structural couplings at the complex-oxide interfaces is a powerful platform for creating new ultrathin layers with electronic and magnetic properties unattainable in the bulk. However, with the capability to design and control the electronic structure of such buried layers and interfaces at a unit-cell level, a new challenge emerges to be able to probe these engineered emergent phenomena with depth-dependent atomic resolution as well as element- and orbital selectivity. Here, we utilize a combination of core-level and valence-band soft x-ray standing-wave photoemission spectroscopy, in conjunction with scanning transmission electron microscopy, to probe the depth-dependent and single-unit-cell resolved electronic structure of an isovalent manganite superlattice [$Eu_{0.7}Sr_{0.3}MnO_3$/$La_{0.7}Sr_{0.3}MnO_3$]×15 wherein the electronic-structural properties are intentionally modulated with depth via engineered oxygen octahedra rotations/tilts and A-site displacements. Our unit-cell resolved measurements reveal significant transformations in the local chemical and electronic valence-band states, which are consistent with the layer-resolved first-principles theoretical calculations, thus opening the door for future depth-resolved studies of a wide variety of heteroengineered material systems.

Rational design and understanding of the electronic properties of new functional materials is a dominant theme in modern experimental and theoretical condensed matter physics and materials science [1-5]. Over the past two decades, epitaxial complex-oxide heterostructuring and interface engineering have emerged as powerful and versatile experimental platforms, enabling the synthesis of electronic, magnetic and structural phases, which are unattainable in bulk crystals or thin films [6-12]. Concurrently, significant strides in the development and refinement of modern materials theories, including various modalities of density functional theory (DFT) [5,13] and dynamical mean-field theory (DMFT) [14,15], have led to the availability of advanced first-principles tools for guiding the synthesis of such heterostructures and interfaces, as well as interpreting experimental results.

Engineering structural couplings at the epitaxial interfaces between perovskite oxides is a promising avenue for atomic-level control of the electronic and magnetic properties in such structures [16-18]. Recent studies of the isovalent $La_{0.7}Sr_{0.3}MnO_3/Eu_{0.7}Sr_{0.3}MnO_3$ (LSMO/ESMO) and $La_{0.5}Sr_{0.5}MnO_3/La_{0.5}Ca_{0.5}MnO_3$ (LSMO/LCMO) superlattices revealed that the emerging highly-localized lattice distortions and non-bulk-like rotations of oxygen octahedra can lead to new electronic and magnetic properties, and provide a way to enhance or suppress functional properties, such as electronic bandwidth and ferromagnetism [19-21]. Furthermore, varying the thicknesses of individual layers within a superlattice above and below the interfacial coupling lengths (2-8 unit cells) adds a powerful control mechanism for tuning these properties at the unit-cell level and as a function of depth. Thus, complex layered oxide structures with custom electronic and magnetic properties, induced by carefully-engineered unit-cell-scale structural modulations, can be constructed via advanced synthesis methods, such as oxide molecular-beam epitaxy (MBE) [22-25].

At the present time, a major challenge in this emergent field is the measurement of highly-depth-dependent electronic properties in such complex layered nanomaterials at the unit-cell scale. The majority of conventional probes of the electronic structure, although extremely useful, provide either surface-sensitive or depth-averaged electronic-structural information (*e.g.* angle-resolved photoemission, scanning-probe spectroscopy, and x-ray absorption). Here, we demonstrate that a combination of core-level and valence-band soft-x-ray standing-wave photoemission spectroscopy (SW-XPS) [26-28] and high-resolution scanning transmission electron microscopy (HRSTEM) [17,20,29] can be utilized to probe the coupling between the electronic and structural properties in an ESMO/LSMO superlattice at the unit-cell level. We extract both core-level and valence-band depth-resolved electronic-structural information from the three individual unit cells of the topmost ESMO layer, which exhibit engineered structural modulations of the A-site-cation positions as well as oxygen-octahedral rotations and tilts. Our experimental results suggest significant local modulations in the valence-band DOS, which exhibit excellent agreement with the first-principles theory and suggest the emergence of a reconstructed ESMO layer at the surface.

**Results**

**Structural depth profiling with high-resolution scanning transmission electron microscopy.** For this study, an epitaxial [3-u.c. LSMO / 3-u.c. ESMO] × 15 superlattice was synthesized on top of a single-crystalline $(La_{0.3}Sr_{0.7})(Al_{0.65}Ta_{0.35})O_3$ (001) substrate by oxide MBE; deposition conditions are reported in Ref. 19. High-resolution scanning transmission electron microscopy (HRSTEM) in conjunction with STEM modeling was used to confirm the presence of structural modulations in the superlattice and to quantify the amplitudes and directions of the A-site cation displacements in each layer (see Methods). Figure 1 provides the summary of the results of this nano-structural analysis as well as the theoretical calculations, starting with the high-angle annular dark-field (HAADF) image of the superlattice along the $[100]_{pc}$ projection (Fig. 1a). The superlattice layering is immediately evident due to the modulations in the brightness of the A-site atomic columns, with the heavier A-site cations (Eu in ESMO) appearing brighter and the lighter ones (La in LSMO) appearing dimmer. The interfaces appear abrupt, with a minimal interfacial intermixing confined to a single unit cell, which is consistent with our prior measurements of similar samples [19,20].

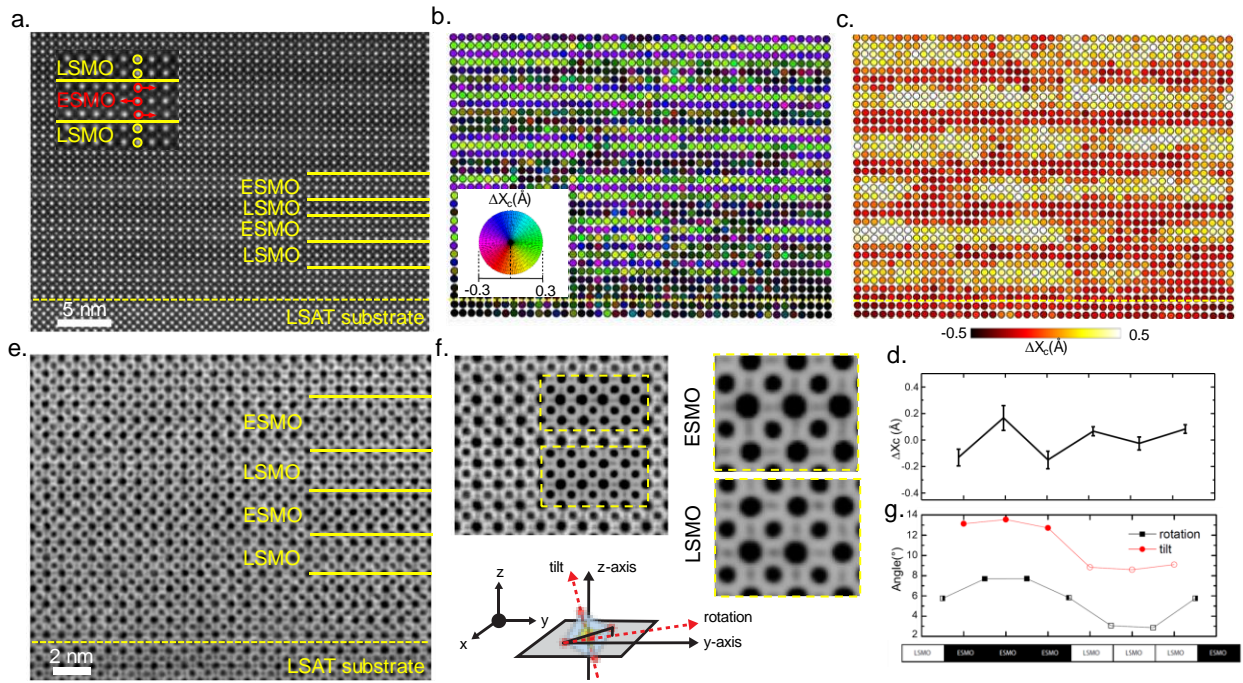

**Figure 1 | Atomic structural modulations: experiment and theory. a.** HRSTEM-HAADF image of the superlattice along the $[100]_{pc}$ projection. The top-left inset shows a magnified image, highlighting local A-site projected displacements. **b.** A-site positions and displacements determined using the HAADF signal. Atomic sites are color-coded according to the amplitude and direction of the displacement with the uncertainties of 0.04 Å and 12.94°, respectively. **c.** A-site displacement gradient map, showing the displacement amplitudes, as referenced to the plane immediately below. **d.** Atomic-plane-averaged A-site displacement amplitudes calculated via DFT+$U$, with the error-bars accounting for the variations within the individual A-site monolayers. **e.** High-resolution ABF image along the $[100]_{pc}$ projection. **f.** LSMO/ESMO supercell built using the averaged data from **e**. and overlaid by the simulated ABF images (yellow dotted boxes). Magnified simulations for ESMO and LSMO are shown in the outsets. **g.** Oxygen octahedral rotation and tilt angles, as defined in the diagram on the left side, calculated via DFT+$U$. The rotation angle is displayed for the equatorial oxygens, while the tilt is shown for the apical oxygens.

The inset in the top-left corner of Fig. 1a shows a magnified view of a typical measured area, containing several atomic layers and highlighting the local A-site projected displacements in the ESMO layer, which are marked with red arrows. Notable zig-zag-like modulations in the projected A-site cation positions are evident in the inset and are quantified for the entire image in Fig. 1b. Here, the atomic sites of the A-site cations are color-coded according to the magnitude and direction of the measured displacement (see figure caption for the measurement uncertainties). The bottom two atomic layers shown in the figure correspond to the substrate, which is used as a zero-displacement reference. Thus, as expected, most of the sites in the substrate appear dark-violet – the color of the center of the HSV wheel in the legend of Fig. 1b. This picture changes abruptly above the substrate, where significant depth-dependent A-site shifts are evident from the color modulations in the first few atomic planes, corresponding to the three-unit-cell-thick ESMO layer. The zig-zag-like pattern, which is shown locally in the inset of Fig. 1a, appears to be a general trend within the ESMO layers, with the alternating amplitudes of approximately +0.3 Å (predominantly green-colored layers) and approximately -0.3 Å (predominantly magenta-colored layers). The presence of A-site displacements is consistent with the structure of bulk ESMO, which crystallizes in the *Pbmn* orthorhombic perovskite variant and exhibits A-site displacements in the plane perpendicular to the in-phase octahedral rotation axis [30]. The modulations are comparatively smaller in the LSMO layers, as evidenced by the broad dark-violet slabs appearing between the ESMO layers. The suppression of the A-site displacements in LSMO is also consistent with its bulk rhombohedral structure, in which the A-site occupies the ideal corner position of the pseudocubic perovskite cell [31]. Fig. 1c shows a differently-color-coded representation of the data, quantifying the absolute changes of the A-site displacements in an atomic plane, as referenced to the plane immediately below. This 'gradient map' is instrumental in emphasizing large displacement gradients in the ESMO layers.

The experimental results for the A-site cation displacements are in full qualitative and close quantitative agreement with the atomic positions predicted by the first-principles DFT+$U$ calculations (see Methods), shown for a typical ESMO/LSMO bilayer within the superlattice. Figure 1d shows the plot of calculated atomic-plane-averaged displacements, with the error-bars accounting for the variations within the individual A-site monolayers. A characteristic zig-zag trend is observed, with prominent modulations in the ESMO layer, which is fully-consistent with the experimental results in Fig. 1b. The calculated displacement magnitudes of approximately +/-0.2 Å (plus the intra-monolayer variations of approximately +/-0.1 Å) are also in good agreement with the experiment (+/-0.3 Å).

Figure 1e shows the high-resolution annual bright field (ABF) image of an area within the same probed region. ABF imaging is sensitive to the oxygen atoms, which do not exhibit sufficient contrast in HAADF due to their low atomic number, as compared to the other elements in the superlattice (Eu, La, Sr, and Mn). Therefore, ABF imaging can be instrumental in detecting and quantifying lattice distortions induced by the changes in the tilt and rotation angles of the oxygen octahedra in perovskite structures [20,32,33]. Such distortions are immediately apparent in Fig. 1e, where oxygen atoms appear as the smallest elongated grey spots in-between the largest black A-site cations. It is also clear, upon a more careful inspection, that the distortions are significantly more pronounced for the ESMO layers. A magnified image of an area containing a typical ESMO and a typical LSMO layers is shown in Fig. 1f. The apparent elongations of the oxygen sites occur due to the variations in the octahedral tilts and rotations (as defined in the schematic diagram below) within the $[100]_{pc}$-projected atomic columns, and appear to be significantly larger in the ESMO layers. This is fully-consistent with the STEM simulation results, overlaid on the experimental data (yellow dotted boxes) and shown in the magnified outsets, as well as the results of the first-principles DFT+$U$ calculations shown in Fig. 1g (see caption for details), which predict a ~4.5° increase in both the tilt and rotation angles relative to LSMO.

In summary, the HRSTEM imaging and simulations confirm the presence of engineered structural modulations in the [ESMO/LSMO]×15 superlattice, in good qualitative and quantitative agreement with the first-principles DFT+$U$ calculations, and consistent with the prior study on similar samples [19,20]. The modulations, manifested as varying A-site cation displacements and oxygen octahedral rotations and tilts, are prominently enhanced in the ESMO layers. In the following, we examine the unit-cell-resolved depth-dependent electronic-structure modulations which accompany these significant lattice distortions in the ESMO layer.

**Electronic and chemical depth profiling with soft x-ray standing-wave photoemission spectroscopy.** In order to selectively probe the depth-resolved electronic structure of each unit cell of the topmost ESMO layer and the ESMO/LSMO interface, we used soft x-ray standing-wave (SW) photoemission spectroscopy (SW-XPS) at the high-resolution ADRESS beamline of the Swiss Light Source [34]. In SW-XPS, depth resolution is accomplished by setting-up an x-ray SW field within a periodic superlattice sample, which in the first-order Bragg reflection acts as a SW generator (see Fig. 2a). The antinodes of the SW (regions of high $E$-field intensity) are then translated vertically through the sample by scanning (rocking) the x-ray incidence angle [26-28]. All measurements were carried out at the photon energy of 833.5 eV, at the onset of the La $M_5$ ($3d_{5/2}$) absorption threshold (characterized in-situ via x-ray absorption spectroscopy), in order to maximize the x-ray optical contrast between ESMO and LSMO, which in-turn lead to the significant enhancement of the SW modulation amplitude [27,35]. The total energy resolution was estimated to be approximately 100 meV, and the sample temperature was set at 30 K.

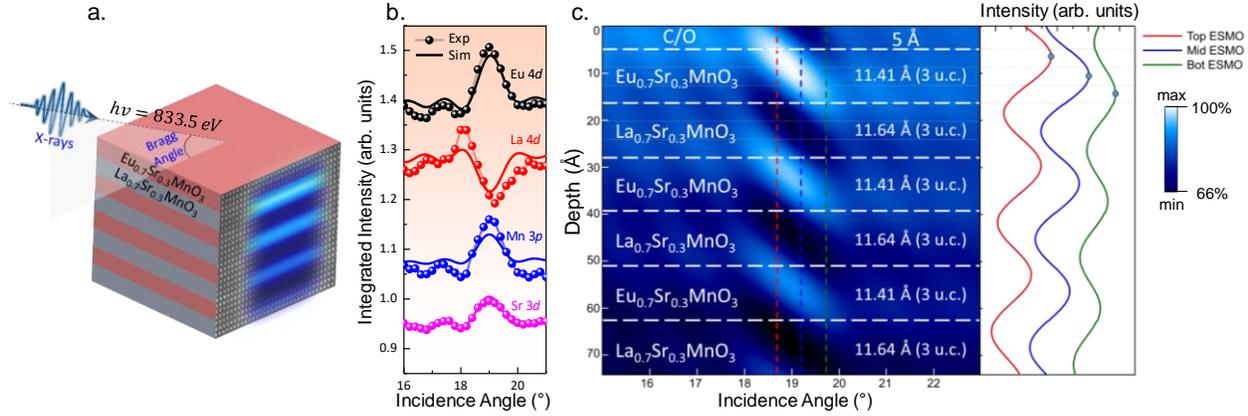

**Figure 2 | SW-XPS experiment and x-ray optical simulations. a.** Schematic diagram of the sample and the experimental geometry, showing the soft x-ray beam, incident at the grazing angle corresponding to the first-order Bragg condition, and the resultant x-ray SW within the superlattice. **b.** The best fits between the experimental and calculated SW rocking-curves for the Eu 4$d$, La 4$d$, Mn 3$p$, and Sr 3$d$ core levels. **c.** The resultant model of the superlattice, which self-consistently describes the shapes and amplitudes of the rocking-curves for every constituent element in the structure (O is shown separately, in Fig. 3). The white-to-blue color scale represents the simulated intensity of the x-ray SW $E$-field ($E^2$) inside the superlattice as a function of depth and grazing incidence angle. The line-cuts and the corresponding $E$-field intensity plots on the right side show that at the grazing incidence angles of 18.7°, 19.2° and 19.8° the SW preferentially highlights the top, middle and bottom unit-cells of ESMO, respectively.

Core-level photoemission intensities were measured in the near-Bragg-angle variable-incidence experimental geometry shown schematically in Fig. 2a. At least one core-level peak from every constituent element of the multilayer was recorded as a function of grazing incidence angle from 16 to 21° (rocking-curve measurement) and fitted using an x-ray optical code which accounts for the multiple reflections at interfaces, differential electronic cross section of each orbital, as well as the elastic attenuation lengths (EAL) within each layer [36]. Only the thicknesses of the ESMO and LSMO layers and the interface roughness (interdiffusion) were allowed to vary in the model. Experimental results for Eu 4*d*, La 4*d*, Mn 3*p*, and Sr 3*d* (circular markers) as well as the best theoretical fits to the data (solid curves) are shown in Fig. 2b, exhibiting good agreement in terms of both amplitudes and relative phases. The La 4*d* and Eu 4*d* rocking-curves exhibit a 180° phase-shift due to the fact that the La and Eu cations reside in different layers and the period of the SW, in the first order approximation, equals to the period of the superlattice [26,36]. The Mn 3*p* and Sr 3*d* photoemission intensities originate from the elements residing in both layers and are thus dominated by the contributions from the top (ESMO) layer, exhibiting similar phase to the Eu 4*d* rocking-curve and suppressed amplitudes, as expected [26,35,37]. It is important to note that, although the entire superlattice, including the substrate and the surface-adsorbed atmospheric contaminant, must be considered by the model, only the topmost layers are actually relevant for our photoemission measurement due to the limited EAL of photoelectrons at 833 eV (~20 Å) [27,35,38,39].

Figure 2c shows a schematic diagram of several topmost layers of the superlattice, obtained using the set of best-fit parameters. The individual thicknesses of the three-unit-cell-thick layers of ESMO (11.41 Å) and LSMO (11.64 Å) are consistent with the unit-cell constants reported in prior studies [19,20,35]. The thickness of the surface-adsorbed atmospheric contaminant (labelled "C/O") is 5 Å, also consistent with prior studies [35,37]. The blue-to-white color contrast in Fig. 2c shows the simulated intensity of the x-ray SW electric field ($E^2$) as a function of the grazing incidence angle (along the horizontal axis). The SW exhibits maximum contrast of approximately 34% in the vicinity of the Bragg condition (~19°). The intensity is maximized in the topmost ESMO layer and exhibits a depth-dependent evolution as a function of the grazing incidence angle, plotted as a series of vertical line-cuts on the right side of the panel. It is evident that at the incidence angles of 18.7°, 19.2° and 19.8°, the peak intensity of the topmost antinode of the SW preferentially highlights the top, middle and bottom ESMO unit cells, respectively. Due to small interfacial intermixing, the bottom unit cell could also be considered an ESMO/LSMO interfacial layer. According to the prior SW studies, the depth-resolution of the SW-XPS in the soft x-ray regime can be approximately estimated as 1/10 of the multilayer period [28,35,37]. For our sample, the resultant estimate of ~2.3 Å is well within the unit-cell limit. Therefore, we can expect to be able to extract unit-cell-resolved depth-dependent information from the top ESMO layer.

This capability becomes clearly evident upon the examination of the O 1$s$ SW rocking-curve, shown as a photoemission intensity (color) map in Fig. 3a. The horizontal axis represents the binding energy, the vertical axis corresponds to the variable grazing incidence angle and is therefore related to the vertical position of the SW within the layer, as discussed above. The plot in Fig. 3a, therefore, contains the depth-resolved information regarding the distribution and evolution of the chemical and electronic states of the oxygen atoms within the probing range of the SW and limited by the EAL of ~20 Å.

Three horizontal line-cuts at 18.7°, 19.2° and 19.8° (see discussion of Fig. 2c above) yield the unit-cell-specific O 1$s$ spectra shown in Fig. 3b. It is important to note that the SW does not exclusively probe any one given unit-cell within a 3 u.c.-thick layer, but rather amplifies the spectral features originating from that unit-cell, according to the depth-dependent $E$-field intensity distribution within the sample. It is, therefore, expected that we should observe a superposition of multiple spectral components originating from various depths, which either grow or decay in intensities as the antinode of the SW propagates vertically through the layer.

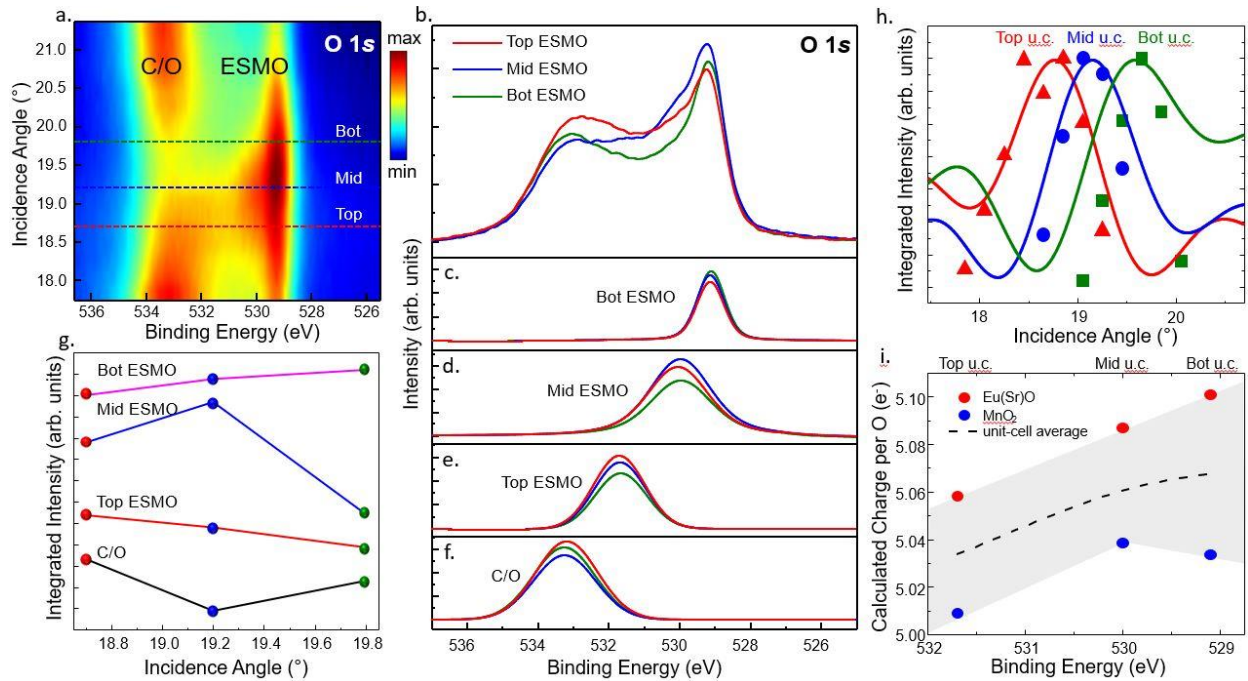

**Figure 3 | Oxygen-derived unit-cell-resolved electronic structure. a.** 2D intensity plot of the depth-dependent evolution of the O 1$s$ core-level, with the three key line-cuts, corresponding to the depths of the bottom (green), middle (blue) and top (red) ESMO unit-cells. **b.** Depth-specific O 1$s$ spectra extracted from the line-cuts in a. **c-f.** Spectral components originating from the bottom (c), middle (d) and top (e) ESMO unit cells, as well as the oxygen-containing surface-adsorbed atmospheric contaminant (f). **g.** Integrated intensities of the unit-cell-specific O 1$s$ spectral components shown in c-f, as a function of the grazing incidence angle. The plots are vertically offset with respect to each other for clarity. **h.** SW rocking-curves of the unit-cell-specific O 1$s$ spectral components (solid symbols) and the rocking-curve simulations for each individual unit-cell comprising the top-most ESMO layer in the superlattice. **i.** Plot of the correlation between the experimental binding-energies of the unit-cell-specific O 1$s$ spectral components and the DFT+$U$-calculated integrated charge on the O atoms in the top three unit cells of ESMO (shown for the Eu(Sr)O and MnO$_2$ planes separately).

These four distinct spectral components, easily identifiable in Fig. 3b, were decoupled via simultaneous fitting of the O 1$s$ spectra with four simple Voight peaks, and plotted separately in Figs. 3c-f for the three probing depths selected by the SW by varying the grazing incidence angle (see Fig. 2c for reference). The integrated intensities for each component are plotted as a function of increasing grazing incidence angle (and therefore increasing probing depth) in Fig. 3g (the curves are offset vertically with respect to each other for clarity).

Each component exhibits a unique angle-dependent behavior. The lowest-binding-energy component (~529.1 eV) exhibits a near-linear growth in intensity with increasing incidence angle, and therefore must originate from the deepest 'bottom' unit cell at the ESMO/LSMO interface. The second component (at $E_B \approx 530.1$ eV) grows in intensity as the SW antinode propagates toward the center of the ESMO layer, and then decays as it approaches the bottom ESMO/LSMO interface. This suggests that it originates from the 'middle' unit cell of the ESMO layer - the SW antinode passes through it, causing an increase in intensity at intermediate angles. The third component (at $E_B \approx 531.7$ eV) continuously decays in intensity with increasing grazing incidence angle and therefore must originate from the 'top' unit cell of the ESMO layer - the SW antinode is continuously moving downward and away from it. Finally, the highest-binding-energy component (at $E_B \approx 533.2$ eV) decays in intensity at intermediate angles but shows a small upturn at the highest angle of 19.8°. Due to its binding energy, this spectral component can be assigned to the oxygen in the surface-adsorbed contaminant [40,41]. The upturn in intensity at 19.8° is caused by another SW antinode grazing the surface of the sample at higher incidence angles, resulting in enhanced photoemission signal from the surface adsorbates. In summary, the unique angle-dependent SW-induced behavior of the distinct spectral components of the O 1*s* spectrum allows for an unambiguous assignment of these components to the distinct layers in the structure. Below, we verify this assignment via rigorous x-ray optical analysis.

In Figure 3h, we plot the experimental SW rocking-curves of the three higher-binding-energy components of the O 1*s* peak (solid markers). It is immediately apparent that the three experimental rocking-curves are shifted with respect to each other in angular position (phase), suggesting a different depth-of-origin (see Fig. 2c) [37,42], which is consistent with our prior analysis, as shown in Figs. 3c-g. The solid curves overlaying the experimental data are the x-ray optical simulations [26] of the rocking-curves for each individual unit-cell comprising the top-most ESMO layer in the superlattice, defined to be 3.803 Å-thick, consistent with the model in Fig. 2c, as well as the unit-cell constants reported in prior studies [19,20,35]. The bottom simulated unit-cell includes the interface with the LSMO underlayer. The agreement between experiment and the simulation is remarkable, in particular, with respect to the shifts in the angular positions of the peaks, which, in turn, correspond to the differences in the depths-of-origin for the maximum photoemission signal. It is important to note that all three experimental peaks occur within the angular range between 18° and 20° and exhibit the lineshape and the phase similar to that of the Eu 4$d$ rocking-curve, shown in Fig. 2b (black spectrum). This serves as an additional verification that all three components originate from the different depths within the ESMO (and not the LSMO or the C/O) layer.

Our rigorous x-ray optical simulations, therefore, confirm the observed eV-scale unit-cell-dependent changes in the binding-energy of the O 1$s$ core-level peak within the 3 u.c.-thick ESMO layer, and thus suggest significant depth-dependent transformations in the chemical/electronic environment around the oxygen atoms within this layer. Such unit-cell-specific variations, undetectable by the conventional depth-averaging and/or surface-sensitive characterization techniques, are not unexpected in view of our HRSTEM results (see Fig. 1), which reveal significant structural modulations within the ESMO layer, consistent with the first-principles DFT+$U$ calculations. Furthermore, symmetry-breaking due to the presence of the surface (as well as strain) may lead to both structural and electronic surface reconstruction phenomena, which could account for the ~0.6 eV increase in the binding energy of the O 1$s$ core-level for the topmost ESMO unit cell (with respect to the unit-cell below).

In order to understand the significant increase in the binding energy of the O 1$s$ core level at the surface, in Fig. 3i we show the results of the DFT+$U$ calculation for the integrated electronic charge on the oxygen atoms for the top three unit-cells of ESMO. It should be noted that only the 2$s$ and 2$p$ orbitals were included in the calculation, with the Wigner radius of integration set to 1 Å, in order to sample the deeper levels, rather than the bonding electrons. The resultant values for the integrated electronic charge exhibit a nearly-linear (inverse) correlation with the O 1$s$ binding energies, with the surface unit-cell exhibiting the lowest charge (~5.034 e$^-$) and the highest binding energy (531.7 eV), as expected from basic considerations [43].

The relatively-large shift in the binding energy of the O 1$s$ core level in the topmost (surface) unit-cell of ESMO suggests the possibility of surface reconstruction/relaxation. We explore this likely scenario below, using depth-resolved SW valence-band photoemission measurements [27,42] in conjunction with the first-principles DFT+$U$ density-of-states (DOS) calculations.

**Depth-resolved valence-band electronic-structure measurements and calculations.** Figure 4a shows the calculated structures of the three topmost unit cells of ESMO (top view). While the bottom and the middle unit cells exhibit structural modulations, which are consistent with our HRSTEM measurements (A-side displacements and oxygen octahedral rotations/tilts), the topmost layer exhibits a new relaxed structure, characterized by the emergence of tilted MnO$_4$ oxygen tetrahedra (with triangular bases), interspersed among the typical MnO$_5$ oxygen square pyramids (surface truncated octahedra). The two above-mentioned Mn-O polyhedra are identified and shown in the outsets of Fig.4a. The change in transition-metal coordination and valence at the surface within our model is due to truncating the crystal. The ordered arrangement we identified results from the ordered A-cations in the simulation cell; the experimental surface geometry, however, may be more complex or exhibit a different ordered arrangement of the tetrahedral and square-pyramidal polyhedra. Below, we demonstrate that such surface-relaxation phenomena, which occur only in the top unit-cell of an epitaxial oxide film (ESMO), can be probed by the depth-resolved SW-XPS of the valence-bands with single-unit-cell resolution.

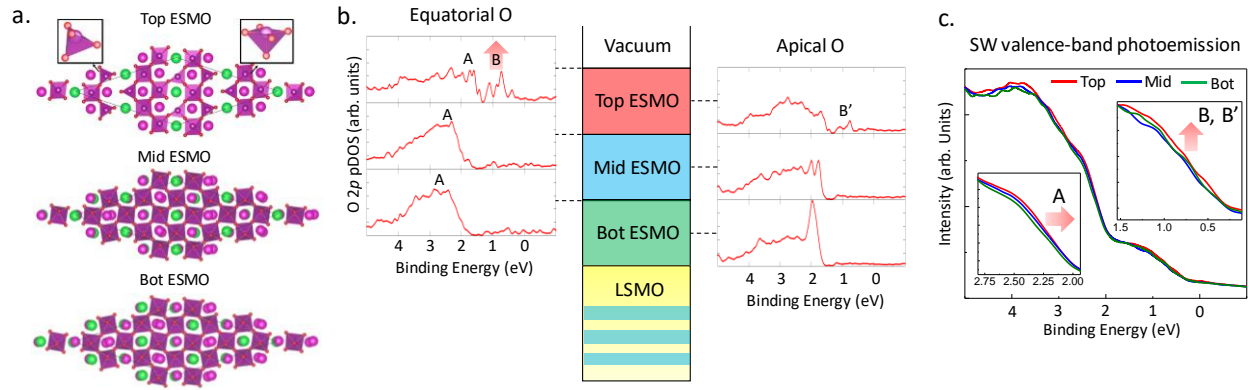

**Figure 4 | Unit-cell-resolved valence-band electronic structure. a.** DFT+$U$ calculations of the atomic structure of the top three unit-cells of ESMO. The layers exhibit structural modulations (in agreement with the HRSTEM measurements), as well the surface-layer reconstruction, characterized by the emergence of the tilted oxygen tetrahedra. **b.** Atomic-plane-resolved O 2$p$-projected pDOS for the atomic planes containing the equatorial and apical oxygen atoms in the topmost ESMO layer. **c.** Unit-cell-resolved SW valence-band photoemission spectra, probing the corresponding depth-resolved changes in the matrix-element-weighted DOS.

Figure 4b shows the effects of the oxygen-mediated surface reconstruction on the layer-resolved valence-bands DOS calculated via DFT+$U$. The region near the Fermi level is dominated by the strongly-hybridized O 2$p$-Mn 3$d$ states. We therefore only show the O 2$p$-projected partial DOS for each atomic plane containing equatorial and apical oxygens. The most significant changes are predicted to occur within the binding-energy window between 0 and 3 eV. In particular, we observe a broadening and a shift to lower binding energy of feature A (at ~2.5 eV), as well as the emergence of a new state at ~1 eV (labeled B and B'), which is particularly strongly-pronounced for the equatorial (surface-like) oxygens (B).

Our unit-cell-resolved experimental SW-XPS valence-band spectra (Fig. 4c) exhibit agreement with the theoretical DOS, both in terms of the energies and the systematic trends in the relative intensities of the relevant features. It is important to note that we expect to see smaller effects in our experimental data (compared to theory), since the SW contrast is estimated to be approximately 34% (see Fig. 2c), which means the unit-cell-dependent changes will ride on a strong depth-averaged background signal. Furthermore, feature B (B') is expected to be prominent in all spectra due to its surface-origin. Nevertheless, we clearly observe a theoretically-predicted shift to lower binding-energy for feature A (at ~2.5 eV, consistent with the calculations). Furthermore, we similarly observe an enhancement in intensity of feature B, B' (at ~1 eV) in the surface (top) ESMO unit-cell.

**Discussion**

In summary, our unit-cell-resolved experimental data for the ESMO/LSMO superlattice, obtained via multiple depth-resolved spectroscopic and microscopic techniques, exhibit excellent agreement with the first-principles layer-resolved DFT+$U$ calculations at several important levels. First, the atomic structure measured via HRSTEM in the bulk of the superlattice is in both qualitative and quantitative agreement with the structure predicted by the theory (see Fig. 1). Second, the depth-dependent shifts in the binding-energy of the O 1$s$ core-level, measured via SW-XPS, exhibit near-linear correlation with the calculated integrated electronic charge on the oxygen atoms for each unit cell of the topmost ESMO layer (see Fig. 3). Third, the depth-dependent SW-XPS of the valence bands, in conjunction with the DOS calculations within the same self-consistent DFT+$U$ model, strongly suggest the emergence of a surface-reconstructed (relaxed)

ESMO layer, characterized by the presence of sites with tetrahedral oxygen coordination (see Fig. 4).

In addition to revealing a new reconstructed surface phase of ESMO, as well as the significant unit-cell-resolved modulations of the core-level and valence-band electronic structure in this transition-metal oxide induced by heterostructuring and strain, these results demonstrate both the power and necessity of depth-resolved x-ray techniques (such as SW-XPS) that are capable of probing buried layers and interfaces and thus go beyond conventional surface-specific or depth-averaging electronic-structure studies.

## Methods

**HRSTEM Measurements and simulations.** Thin TEM lamellae were prepared by using a FEI Strata 235 dual beam Focused Ion Beam followed by Ar polishing at 500 eV and final cleaning at 200 eV, using Fischione NanoMill.

HRSTEM measurements were performed using an aberration-corrected TEAM I FEI Titan-based TEAM I microscope at the National Center for Electron Microscopy Facility of the Molecular Foundry (LBNL), operated at 300 kV with probe current kept at approximately 70 pA during data acquisition. HAADF and ABF images were acquired using a convergence semi-angle of 30 mrad and 17.2 mrad, respectively. Collection angle were set to 40-240 mrad range for HAADF and 6-30 for ABF images. Raster distortion was minimized using a pair of HRSTEM images of the same region taken with orthogonal scan orientations to minimize the slow scan axis errors [44]. Post-processing of HAADF images for measuring the A-site displacements were performed using custom MATLAB codes. A 2D gaussian function was used to fit the A-site positions and then compare to a mean unit cell lattice extracted by averaging out a region comprising the substrate $(La_{0.3}Sr_{0.7})(Al_{0.65}Ta_{0.35})O_3$.

STEM image simulations of the crystal model obtained by DFT were carried out using multislice methods of Kirkland [45] under PRISMatic environment [46,47] applying sixteen frozen phonons and the experimental parameters (voltage and probe semi-convergence angle).

**First-principles calculations.** DFT calculations were performed using Vienna Ab-initio Simulation Package (VASP) [48,49] to provide insights on structures for both the internal and surface layers of the ESMO/LSMO superlattice. The revised Perdew-Burke-Ernzerhof (PBE) functional for solids (PBEsol) [50] were used in our calculation. The projector augmented wave (PAW) method [51] was used to treat the core and valence electrons using the following electronic configurations: $6s^25s^25p^65d^1$ for La, $6s^25s^25p^65d^1$ for Eu, $4s^22p^65s^2$ for Sr, $3p^64s^23d^7$ for Mn, and $2s^22p^4$ for O. The electron correlation effects of Mn $3d$ states were considered by the inclusion of the Hubbard $U$ (PBE+$U$) [52], with $U_{Mn} = 3$ eV.

In order to accommodate the 1/3 doping level of our system, as well as the 3×3 superlattice period, a $\sqrt{10} \times \sqrt{10} \times 3$ supercell (with respect to the orthorhombic cell) was used. A 25 Å-thick vacuum layer was used to simulate the surface. The atomic positions were relaxed until the force on each atom are less than 0.01 eV/ Å. Only the Γ-point was sampled for this large system during structure optimization, while a $3 \times 3 \times 2$ Γ-centered Monkhorst-Pack $k$-point mesh was used for calculating the electron DOS. Integrations are performed using Gaussian smearing with a width of 20 meV.

## Acknowledgements

A.X.G., R.U.C., W.Y. and A.A. acknowledge support from the U.S. Army Research Office, under Grant No. W911NF-15-1-0181. A.X.G. also acknowledges support from the US Department of Energy, Office of Science, Office of Basic Energy Sciences, Materials Sciences and Engineering Division under award number DE-SC0019297 during the writing of this paper. E.J.M. and S.J.M. acknowledge support from the U.S. Army Research Office, under Grant No. W911NF-15-1-0133. M.G. and J.M.R. were supported by the U.S. Department of Energy (DOE) under Grant No. DE-SC0012375. J.C. and R.M.S.dR. acknowledge additional support from the U.S. Department of Energy Early Career Research Program. M.-A.H. was supported by the Swiss Excellence Scholarship grant ESKAS-no. 2015.0257. DFT calculations were performed using the CARBON Cluster at Argonne National Laboratory (DOE-BES, Contract No. DE-AC02-06CH11357). Work at the Molecular Foundry user facilities was supported by the Office of Science, Office of Basic Energy Sciences, of the U.S. Department of Energy under Contract No. DE-AC02-05CH11231.

## Author Contributions

W.Y., R.U.C. and A.A. carried out the experiments and analyzed the data, in collaboration with V.N.S., M.-A.H, J.C., and R.dR., and under the supervision of A.X.G. Samples were grown by E.J.M., under the supervision of S.J.M. Density functional theory calculations were carried out by M.G., under the supervision of J.M.R. All the co-authors contributed to writing and editing the paper.

**Competing interests:** The authors declare no competing interests.